\documentclass[showpacs,aps,prr,floats,twocolumn, notitlepage,superscriptaddress,nofootinbib,longbibliography]{revtex4-2}
 \pdfoutput=1

\usepackage{graphicx}
\usepackage{float}
\usepackage{braket}
\usepackage{color}
\usepackage{epstopdf}
\usepackage{mathtools}
\usepackage{amsmath}
\usepackage{amssymb}
\usepackage[symbol]{footmisc}

\usepackage[caption=false]{subfig}

\usepackage{hyperref}
  \hypersetup{
      colorlinks=true,
      linkcolor=blue,
      filecolor=blue,
      citecolor = blue,      
      urlcolor=black,
      }

\usepackage{float}

\newcommand{\comment}[1]{}
\newcommand{\remove}[1]{}

\newcommand{\blue}[1]{{\color{black}#1}}

\makeatletter
\newcommand\blfootnote[1]{%
  \begingroup
  \renewcommand\thefootnote{}\footnote{#1}%
  \addtocounter{footnote}{-1}%
  \endgroup
}

\begin{document}

\makeatother
\makeatletter \def\@fnsymbol#1{\ensuremath{\ifcase#1\or *\or \dagger\or \ddagger\or
   \mathsection\or \mathparagraph\or \|\or **\or \dagger\dagger
   \or \ddagger\ddagger \else\@ctrerr\fi}}
\renewcommand*{\thefootnote}{\fnsymbol{footnote}}
\makeatother
\title{High-resolution spectroscopy of a quantum dot driven bichromatically by two strong \blue{coherent} fields 
}

\author{Chris Gustin$^\dagger$ }
\thanks{These authors contributed equally to this manuscript}
\affiliation{Department of Applied Physics, Stanford University, Stanford, California 94305, USA}
\affiliation{Department of Physics, Engineering Physics, and Astronomy, Queen's University, Kingston, Ontario K7L 3N6, Canada}
\author{Lukas Hanschke$^\dagger$}
\thanks{These authors contributed equally to this manuscript}
\affiliation{Walter Schottky Institut and Department of Electrical and Computer Engineering, Technische Universit\"at M\"unchen, 85748 Garching, Germany}
\affiliation{Munich Center for Quantum Science and Technology (MCQST), Schellingstr. 4, 80799 Munich, Germany}
\author{Katarina Boos}
\thanks{These authors contributed equally to this manuscript}
\affiliation{Walter Schottky Institut and Department of Electrical and Computer Engineering, Technische Universit\"at M\"unchen, 85748 Garching, Germany}
\affiliation{Munich Center for Quantum Science and Technology (MCQST), Schellingstr. 4, 80799 Munich, Germany}
\author{Jonathan R. A. M\"uller$^{\ddagger}$}
\affiliation{Walter Schottky Institut and Physik Department, Technische Universit\"at M\"unchen, 85748 Garching, Germany}
\author{Malte Kremser}
\affiliation{Walter Schottky Institut and Physik Department, Technische Universit\"at M\"unchen, 85748 Garching, Germany}
\affiliation{Munich Center for Quantum Science and Technology (MCQST), Schellingstr. 4, 80799 Munich, Germany}
\author{Jonathan J. Finley}
\affiliation{Walter Schottky Institut and Physik Department, Technische Universit\"at M\"unchen, 85748 Garching, Germany}
\affiliation{Munich Center for Quantum Science and Technology (MCQST), Schellingstr. 4, 80799 Munich, Germany}
\author{Stephen Hughes}
\affiliation{Department of Physics, Engineering Physics, and Astronomy, Queen's University, Kingston, Ontario K7L 3N6, Canada}
\author{Kai M\"uller}
\affiliation{Walter Schottky Institut and Department of Electrical and Computer Engineering, Technische Universit\"at M\"unchen, 85748 Garching, Germany}
\affiliation{Munich Center for Quantum Science and Technology (MCQST), Schellingstr. 4, 80799 Munich, Germany}
\begin{abstract}
We present spectroscopic experiments and theory of a quantum dot driven bichromatically by two strong coherent lasers.
In particular, we explore the regime where the drive strengths are substantial enough to merit a general non-perturbative analysis, resulting in a rich higher-order Floquet dressed-state energy structure.
We \blue{show} 
 high resolution spectroscopy measurements with a variety of laser detunings
 performed on a single InGaAs quantum dot, with the resulting features well explained
 with a time-dependent quantum master equation
 and Floquet analysis.
Notably, driving the quantum dot resonance and one of the subsequent Mollow triplet sidepeaks, we  observe the disappearance and subsequent reappearance of the central transition \blue{and transition resonant with detuned-laser} at high detuned-laser pump strengths and additional higher-order effects, e.g. emission triplets at higher harmonics and signatures of higher order Floquet states. For a similar excitation condition but with an off-resonant primary laser, we observe similar spectral features but with an enhanced inherent spectral asymmetry.%
\end{abstract}

\date{\today}
\maketitle

\section{Introduction}

\blfootnote{$^\dagger$ \text{cgustin@stanford.edu, lukas.hanschke@wsi.tum.de} \\ $^{\ddagger}$ \text{present address: Toshiba Research Europe Limited, Cambridge} \\ \text{Research Laboratory, Cambridge CB4 0GZ, UK}}
Semiconductor quantum dots (QDs) provide an excellent solid-state platform for the coherent control of quantum light-matter interactions. In particular,  optically-active excitons (electron-hole pairs) can behave as mesoscopic two-level systems, allowing for controlled emission of radiation for various forms of quantum information processing protocols, including the generation of single photons~\cite{grange17,ding16,senellart17,somaschi16,schweickert18,thomas20,dusanowski19,liu18} and entangled photon pairs~\cite{huber17,huber18,chen18,huber14,olbrich17,liu19}. Coherent control of quantum systems via continuous wave or pulsed
lasers allows for additional tailoring of the emitted photon spectrum by enabling engineered quantum evolution under strong field interaction, which can manifest in strong-field observables such as
the Mollow triplet, Ramsey interference, and Rabi oscillations~\cite{mollow69,muller07,dory16,jayakumar13,xu07,ulhaq12}.

To provide further control of the dressed-state spectrum, two or more coherent drives can be introduced into the excitation scheme, with 
potential applications including enhanced phonon reservoir squeezing~\cite{gao16}, suppression of the resonant spontaneous emission spectral line (which can overlap with the scattered laser)~\cite{he15}, spectral line narrowing~\cite{li99}, and gain without population inversion~\cite{peiris14}. Periodic driving of quantum systems also can suppress decoherence via continuous dynamical decoupling and the coherent destruction of tunneling~\cite{grossmann91,fonseca05,xu12}. The specific case of two-level systems driven by two coherent drives of differing frequencies (bichromatic), initially motivated by amplitude-modulated driving, has been theoretically studied (e.g., see Refs.~\onlinecite{agarwal91,ficek93,papageorge12,ficek96,blind80}), as well as experimentally using QDs~\cite{peiris14,he15,maragkou13}, atoms~\cite{zhu90,yu97}, and superconducting qubits~\cite{pan17}. Experiments on bichromatically driven QDs have studied certain regimes of two-color excitation (including ``doubly dressed" states), and have revealed an interference-based suppression of the spectral emission line resonant with the exciton frequency, when driven with a strong resonant laser and a second laser detuned to one of the sidebands of the resultant Mollow triplet, as well as a multiphoton AC stark shift of subharmonic resonances~\cite{he15,rudolph98}. 

In this work, we 
explore the regime where both laser drives are strong enough to create a significant component of the Hamiltonian which is, even in any rotating frame, periodic in time, resulting in a rich Floquet dressed-state energy structure, where a general non-perturbative analysis is warranted.  We provide high resolution spectroscopy measurements, the results of which are well replicated with a time-dependent master equation approach, non-perturbative in the coherent drive strengths with respect to the periodicity of the Hamiltonian, which is conceptually straightforward and can easily be generalized to include,  e.g.,  exciton-phonon interactions. We also elucidate how the manifold of bichromatically dressed states which arise from the time-periodic Hamiltonian can be calculated to arbitrary order in harmonic expansion with a Floquet approach, and show an excellent agreement in the transitions it predicts with the full calculations and experiments. 

Our experiments reveal,  for the first time  to our knowledge: \blue{(i)} the higher-order effect of the resonant spectral line re-emergence at high detuned-laser pump strengths~\cite{ficek96}; \blue{ and (ii) the disappearance and subsequent reappearance of the spectral line resonant with the secondary laser with increasing secondary laser power, both for the specific excitation scenario of a resonant primary laser and a secondary laser detuned to a Mollow sidepeak of the primary laser}. We also confirm suppression of spectral lines due to quantum interference and coupling to subharmonic resonances which has already been \blue{observed} by He {\it et al.}~\cite{he15} and observe additional features of the fluorescence spectrum, including the formation of triplets centered at higher multiples of the Rabi energy of the driving laser. Furthermore, by driving the QD under a similar excitation scenario but with an off-resonant primary laser, we observe similar spectral features as the resonant case but with heightened inherent spectral asymmetry, in contrast to the (pure dephasing free) monochromatic case~\cite{gustin18}.

The rest of this paper is organized as follows: in Sec.~\ref{sec:th} we outline the theory of our time-dependent master equation approach. We describe our high resolution spectroscopy measurements in Sec.~\ref{sec:exp}, and detail the experimental and theoretical results in Sec.~\ref{sec:re}. In Sec.~\ref{sec:c} we conclude. We also include \blue{three} appendices: in Appendix~\ref{A:A}, \blue{we describe our semi-analytical Floquet method for calculating the position of potential spectral lines, as well as include an energy level diagram and perturbative analytical calculations for the specific case of a weak secondary drive dressing a sidepeak of a Mollow triplet created by a strong primary laser}; \blue{in Appendix~\ref{A:B}, we present data of experimental characterization of the QD};  and in Appendix~\ref{A:C}, we provide single coherent drive experiments and extract from them an estimate of the phonon coupling strength, which allow us to verify that electron-phonon coupling is qualitatively insignificant in the bichromatic driving regimes studied in the main text.

\section{Quantum master equation
and incoherent spectra}\label{sec:th}

In this section we provide the theory of
bichromatic driving in the
strong field  regime, using
a quantum master equation 
with a time-dependent drive.

We model a bichromatically driven QD as a two-level system, with ground $\ket{g}$ and exciton $\ket{x}$ states. The QD is coherently driven by two lasers at frequencies $\omega_1$ and $\omega_2$, treated semiclassically, with Rabi energies $\Omega_1$ and $\Omega_2$, respectively. The first laser is detuned from the exciton frequency ($\omega_x$) by $\Delta_1 = \omega_x - \omega_1$, and the second laser is detuned by $\Delta_2 = \omega_x - \omega_2$. After making the rotating-wave approximation with respect to the dipole-field interaction term, our system Hamiltonian is periodic with frequency $\Delta=\Delta_1-\Delta_2 = \omega_2 - \omega_1$, and in a frame rotating at $\omega_1$, is (letting $\hbar=1$ throughout)
\begin{equation}\label{hamiltonian}
H(t) = \Delta_1 \sigma^+ \sigma^- + \frac{1}{2}\Big[(\Omega_1+\Omega_2 e^{-i\Delta t})\sigma^+ + \text{H.c.}\Big],
\end{equation}
with the  Pauli operators $\sigma^- = \ket{g}\bra{x}$, $\sigma^+ = \ket{x}\bra{g}$. For QDs, our two-level approximation is justified if we assume the detunings to be small enough as to be far off from any resonances involving multi-exciton states -- a requirement easily satisfied here. For convenience, we also  define the ratio of Rabi energies as $\alpha_c = \Omega_2/\Omega_1$.

We incorporate spontaneous emission at rate $\gamma$ into the model with an open-system Lindblad master equation for the density operator $\rho$:
\begin{equation}\label{eq:me}
\frac{\rm{d}\rho}{\rm{dt}} = -i[H(t),\rho] + \frac{\gamma}{2}\mathcal{L}[\sigma^-]\rho + \frac{\gamma'}{2}\mathcal{L}[\sigma^+\sigma^-]\rho,
\end{equation}
where $\mathcal{L}[A]\rho = 2A\rho A^{\dagger}-A^{\dagger}A\rho -\rho A^{\dagger}A$;  we have also included a phenomenological pure dephasing rate $\gamma'$, to capture linewidth broadening, as well as effects including charge noise~\cite{kuhlmann13}, and, notably, electron-phonon coupling. While electron-phonon scattering has important effects on the dynamics of optically excited QDs~\cite{nazir16,ramsay10,krummheuer02,besombes01,forstner03,PhysRevB.80.201311, PhysRevLett.104.157401,roy11,ulhaq13,PhysRevB.80.201311,PhysRevB.92.205406, PhysRevX.5.031006}, for the drive strengths and detunings considered in this work, these effects are small and can be accurately approximated at low temperatures as a pure dephasing rate. Considering the simple case of a single laser drive on resonance with the exciton, this rate is approximately~\cite{nazir08},
\begin{equation}\label{eq:naz}
    \gamma_{\rm{ph}}' \approx \pi k_B T \alpha \Omega^2,
\end{equation}
where $\alpha$ is the phonon coupling strength, $k_B$ is Boltzmann's constant, and $T$ is the temperature. Even at the relatively high drive strength of $\Omega =100 \ \mu \rm{eV}$, for a phonon coupling rate of $\alpha \lesssim 0.1 \ \rm{ps}^2$ (which is extracted from single drive measurements as shown in Appendix~\ref{A:B}), and temperature $T = 4.2 \ \rm{K}$, this pure dephasing rate $\gamma_{\rm{ph}}' \lesssim 2.7 \ \mu \rm{eV}$ is similar to what could be expected from other background dephasing sources, and thus we can neglect phonon coupling beyond this phenomenological treatment for this work (which we have verified numerically by comparing with a full time-dependent polaron-transform model~\cite{mccutcheon10,2gustin18}). \blue{Similar observations have also been made recently in studying a QD driven monochromatically with two distinguishable drives~\cite{macias20}.}

We  calculate the emitted \emph{incoherent} resonance fluorescence spectrum\footnote{\blue{the coherent part of the spectrum corresponds formally to Dirac delta functions at the laser frequencies, as well is removed from the experimental spectra from polarization filtering; thus we do not include it}}
from the QD in terms of the two-time correlation function:
\begin{equation}\label{eq:s}
S_{\rm{i}}(\omega) = \rm{Re}\Big[\!\!\int_0^\infty \!\! d\tau  e^{i(\omega-\omega_1)\tau} \!\! \int_0^{\infty} \!\! dt \langle \sigma_{\delta}^+(t)\sigma_{\delta}^-(t+\tau)\rangle \Big],
\end{equation}
where $\sigma_{\delta}^{\pm} = \sigma^{\pm} - \langle \sigma^{\pm}\rangle$. As the bichromatically driven system dynamics continually oscillate, the two-time correlation function must be time-averaged over $t$, for at least the periodicity of the Hamiltonian, $2\pi/|\Delta|$. Furthermore, this should be done in the steady-state condition, namely once any transient phenomena have decayed to zero (i.e., for $t \gg 1/\gamma$). Also note that the integration period of the $\tau$-integral is determined not by the periodicity of the Hamiltonian, but by the decay time of the two-level system. To make this calculation more manageable and/or obtain analytical results, incoherent spectra for bichromatically driven systems have historically typically been calculated using Floquet expansions~\cite{agarwal91,ficek93,peiris14,he15,maragkou13}; however, we calculate the full correlation function from the master equation solution via the quantum regression theorem, and average directly over $t$, which is more computationally intensive but has the benefit of simplicity and generalizability and avoids some of the subtleties involved with averaging over expansions of correlation functions~\cite{aronstein02}. 

\blue{In addition to the full numerical calculation of the resonance fluorescence spectrum with the quantum master equation, it is also useful to extract the doubly dressed states of the system (the frequencies of which determine the spectral lines seen in the spectrum) by calculation of the Floquet energies of the periodic Hamiltonian~\eqref{hamiltonian}. In Appendix~\ref{A:A}, we show how to calculate the Floquet energy states for the system, and use them to derive analytical expressions for the spectral resonances for the specific case of a weak laser dressing a sidepeak of a Mollow triplet created by a strong primary laser, as well as show a simplified energy level diagram for this setup.}

\section{Experiments}\label{sec:exp}

Experimentally, we \blue{perform} measurements on InGaAs QDs grown via molecular beam epitaxy with the Stranski-Krastanov mode in a GaAs matrix. We include 17 pairs of alternating GaAs/AlAs layers forming a distributed Bragg reflector below the QD layer to increase the extraction efficiency of the emitted photons. An embedded $n$-doped GaAs layer $35$~nm below the quantum dots forms a Schottky diode together with a semi-transparent $5$~nm thick titanium layer evaporated on the surface of the sample. By applying an external voltage, the resulting electric field in the vicinity of the QDs reduces the charge noise. Furthermore, it allows us to deterministically charge the QD and tune the transition of interest in perfect resonance with the excitation laser via the quantum-confined Stark effect~\cite{Warburton2000,Fry2000}. \blue{Further details are given in Appendix~\ref{A:B}.}

The measurements are performed at $4.2 \ \rm{K}$ in a dip stick setup with a confocal microscope, while cross-polarized filtering allows resonant excitation by suppressing the scattered laser in the detection path~\cite{Kuhlmann2013}.
To avoid the influence of higher excited states \blue{and have a clean two-level system as a basis}, we investigate the negatively charged exciton transition which lacks a fine structure splitting compared to the neutral exciton transition~\cite{Bayer2002}, while higher excited states involve the $p$-shell states which are several tens of meV detuned.
For the high resolution spectroscopy of the dressed states, we employ a scanning Fabry-P\'erot cavity with a free spectral range of $30 \ \rm{GHz}$ ($124 \ \mu$eV) and a resolution of $300 \ \rm{MHz}$ ($1.24 \ \mu$eV)
where the transmitted signal is recorded with an APD. The exciton resonance of the investigated QD is at $\omega_x = 1362.04 \ \text{meV}$, with a lifetime of $455$ ps (decay rate $1.44 \ \mu$eV) \blue{(see Appendix~\ref{A:B})}.

\section{Results}\label{sec:re}

\begin{figure}[htb]
        \includegraphics[width = 0.9\columnwidth]{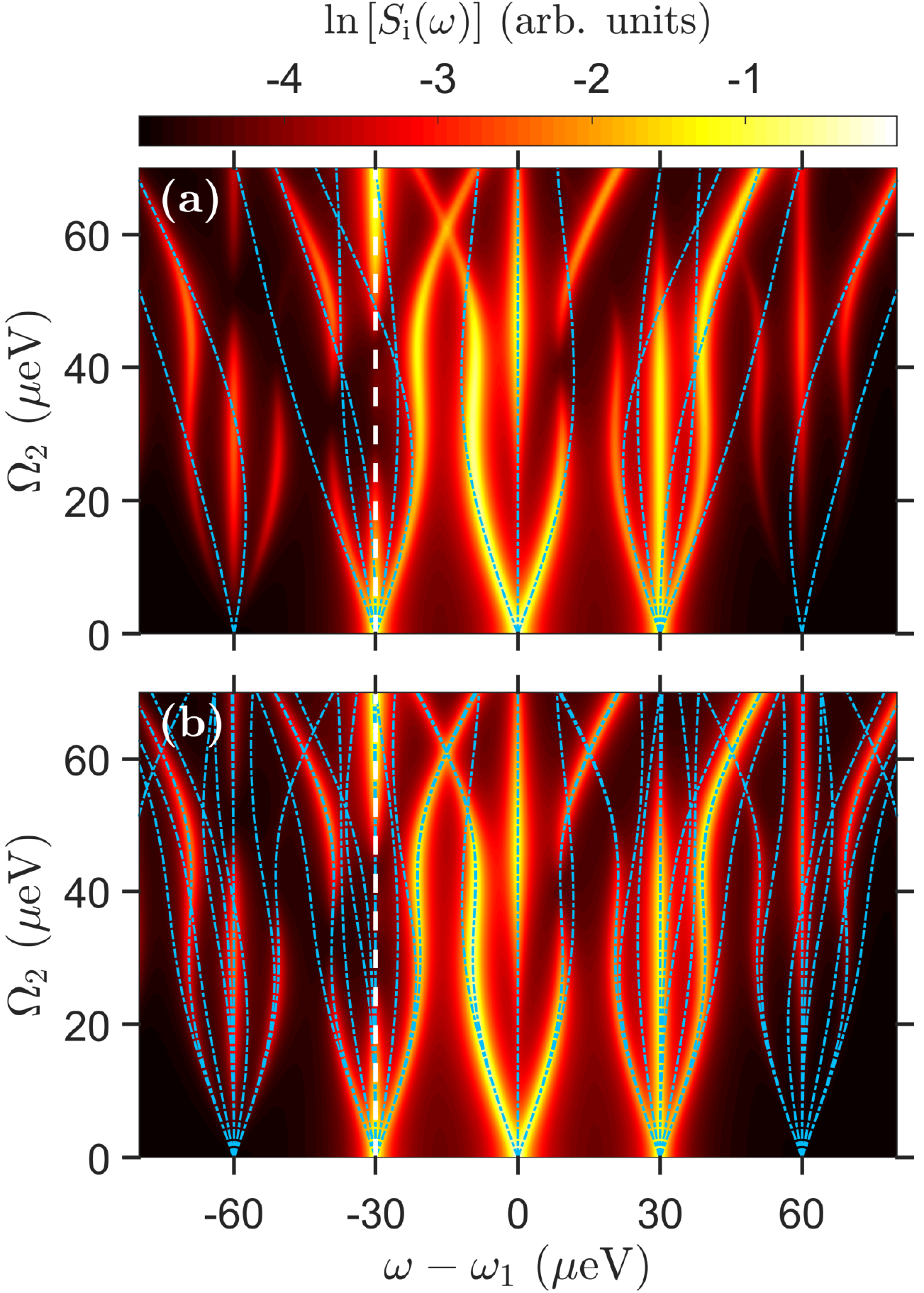}
        \caption{ \small Theoretical log-scale spectrum  \blue{calculated with the master equation (Eq.~\eqref{eq:me})}, with $\gamma = \gamma' = 1 \ \mu \rm{eV}$, $\Omega_1 = 30 \ \mu \rm{eV}$, $\Delta_1 = 0$, $\Delta_2 = 30 \ \mu \rm{eV}$. Floquet transitions up to order (a) $N=1$ and (b) $N = 3$ (the minimum number to see complete agreement with the observed spectral lines) \blue{are shown as dash-dotted lines}. The location of the second laser at $\omega_2$ is shown as a white vertical dashed line.}\label{fig1}
\end{figure}
\begin{figure}[htb]
        \includegraphics[width = 1\columnwidth]{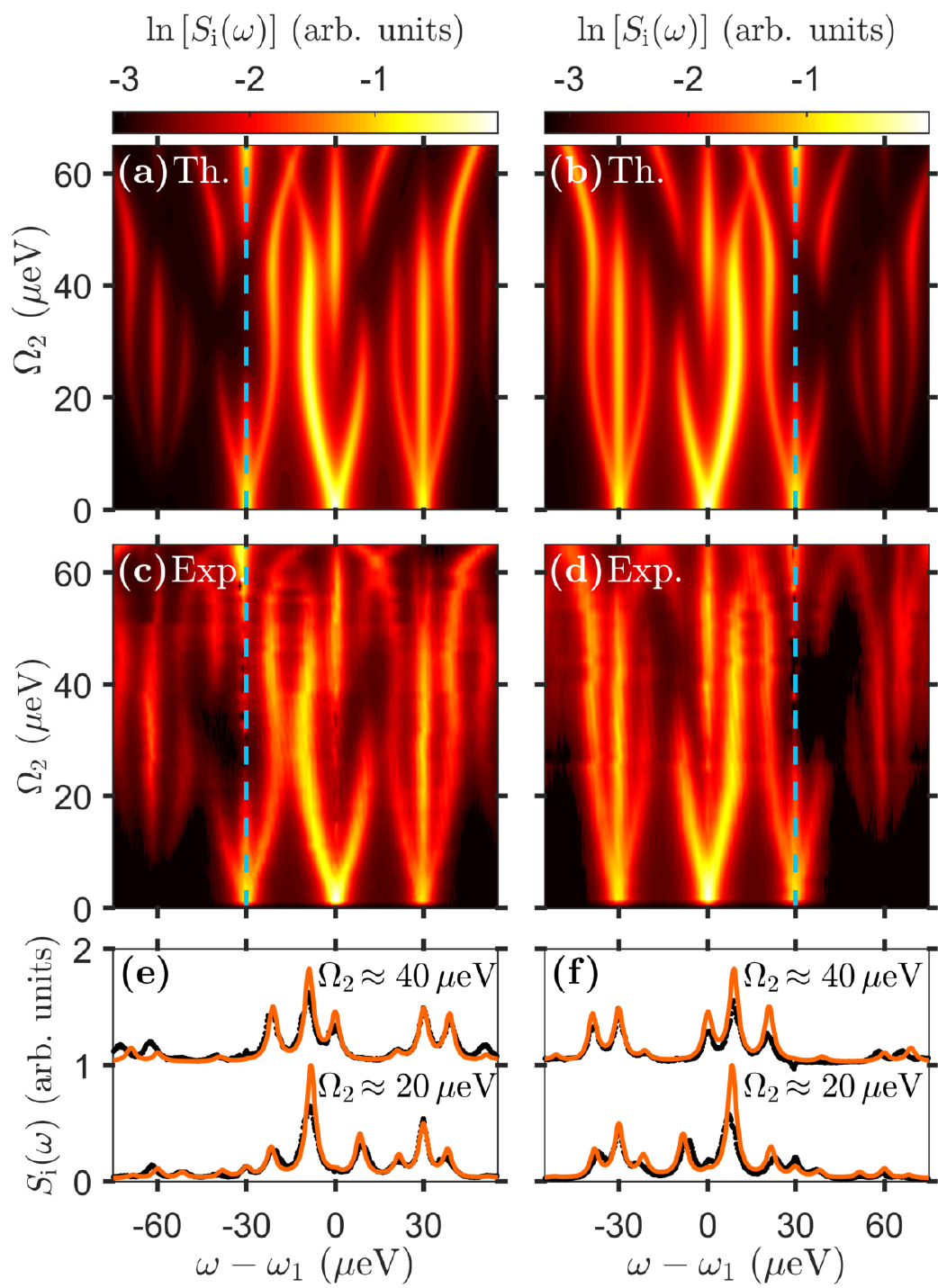}
    \caption{ \small Emission spectrum of a  QD dressed by a resonant ($\Delta_1 = 0$) laser with drive strength $\Omega_1= 30 \ \mu \text{eV}$, as well as a second laser with detuning $\Delta_2 = -\Delta = 30 \ \mu \text{eV}$ for (a,c), and $\Delta_2 = - 30 \ \mu \text{eV}$ for (b,d), and varying drive strength $\Omega_2$. (a,b) show the theoretical calculation with $\gamma = 1.66 \ \mu \text{eV}$ and $\gamma' = 2 \ \mu \text{eV}$, and (c,d) are experimental data. The location of the second laser at $\omega_2$ is shown as a blue vertical  dashed line. (e) and (f) give the data (black) and simulation (orange) for specific values of $\Omega_2$ for (a,c) and (b,d), respectively. At the far ends of the spectra, the Fabry-P{\'e}rot setup leads to a replication of spectral lines separated by the free spectral range.}\label{fig2}
\end{figure}

In Fig.~\ref{fig1}, we plot the theoretically calculated emission spectrum from the bichromatically driven QD, where one laser is held fixed on resonance (with the QD exciton), and the other is detuned to the frequency of the lower energy peak of the resulting Mollow triplet. We also show here the transitions predicted by the Floquet theory overlayed on top, showing excellent agreement with the full numerical calculations of the master equation. The potential Floquet transitions are shown for both $N = 1$ and $N=3$ (which is the lowest integer required to see full agreement with the exact numerical solution), highlighting the higher-order perturbative nature of this pumping regime with respect to $\alpha_c = \Omega_2/\Omega_1$, as well as the rich complexity of the Floquet eigenstructure. This excitation condition, for small $\alpha_c$, gives rise to ``doubly-dressed states''; for low values of $\alpha_c$, each of the Mollow triplet peaks at $\omega_1$, $\omega_1  \pm \Omega_1$ are split by approximately $\pm \Omega_2/2$, creating eight total peaks (the center line being suppressed)~\cite{yu97}. As $\Omega_2$ is increased with $\Omega_1$ held fixed, the center transition line disappears due to destructive interference from transition amplitudes~\cite{he15} and subsequently reappears at higher secondary laser strengths with transition probability having a leading term fourth order in $\alpha_c$~\cite{ficek96}; \blue{furthermore, the transition at $\omega = \omega_2$ also disappears and reappears as $\alpha_c$ is increased}. 


Additionally, higher order   effects (in $\alpha_c$) lead to additional triplets forming at integer multiples of $\Omega_1$ from the center exciton frequency~\cite{ficek96}. Specifically, the leading order transition probabilities for transitions occurring at triplets centered at $\omega = \omega_1 \pm n \Omega_1$, where $n$ is an integer greater than $1$, scale with $\alpha_c^{2(n-1)}$~\cite{ficek96}. This can be understood as a nonlinear multiphoton effect \emph{between} the two detuned driving fields and the exciton. In contrast to analytical calculations perturbative with respect to $\alpha_c$~\cite{ficek96}, which predict a symmetric spectrum about $\omega = \omega_1$, our results show the inherent asymmetry in the spectrum, even for $\alpha_c <1$, and we note this asymmetry persists even if we take $\gamma' = 0$. For $\alpha_c \gg 1$, the structure of this excitation condition changes; it is preferable to not consider this system as a second laser doubly-dressing a Mollow sidepeak, but rather as an off-resonant Mollow triplet centered at $\omega=\omega_2$, dressed by a weaker resonant laser.

\begin{figure}[thb]
        \includegraphics[width = 1\columnwidth]{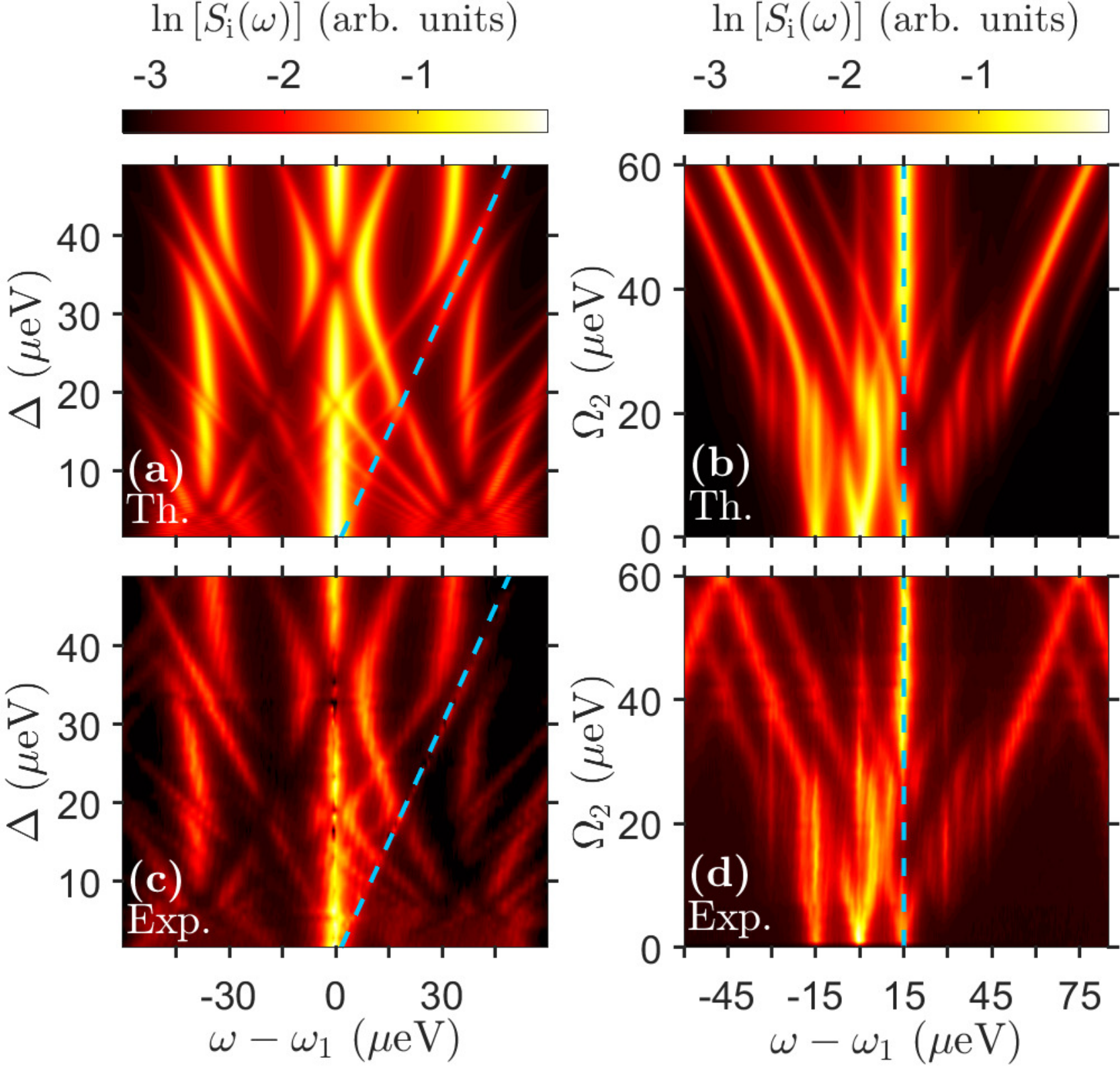}
    \caption{ \small (a,c) Theoretical and experimental log-scale spectra of a QD dressed by a resonant ($\Delta_1 = 0$) laser with fixed drive strengths (for the simulated spectra, $\Omega_1= 35 \ \mu \text{eV}$ and $\Omega_2  = 15 \ \mu \rm{eV}$), where the detuning of the second laser $\Delta = -\Delta_2 = \omega_2 - \omega_1$ is varied. (b,d) Theoretical and experimental spectra of a QD dressed by resonant drive $\Omega_1 = 15 \ \mu \rm{eV}$, with the second laser detuned $\Delta = 15  \ \mu \rm{eV}$ and allowed to vary in strength, revealing a Mollow triplet forming centered at $\omega-\omega_2$ for $\Omega_2 \gg \Omega_1$. For both theoretical calculations, $\gamma = 1.66 \ \mu \rm{eV}$, and $\gamma' = 2 \ \mu \rm{eV}$. The location of the second laser at $\omega_2$ is shown as a thin blue dashed line. In (d), the Fabry-P{\'e}rot setup leads to a replication of spectral lines separated by the free spectral range, seen in the regions $\omega - \omega_1  < -45 \ \mu \text{eV}$, and $\omega - \omega_1  > 75 \ \mu \text{eV}$.}\label{fig3}
\end{figure}

\begin{figure}[hbt]
        \includegraphics[width = 0.9\columnwidth]{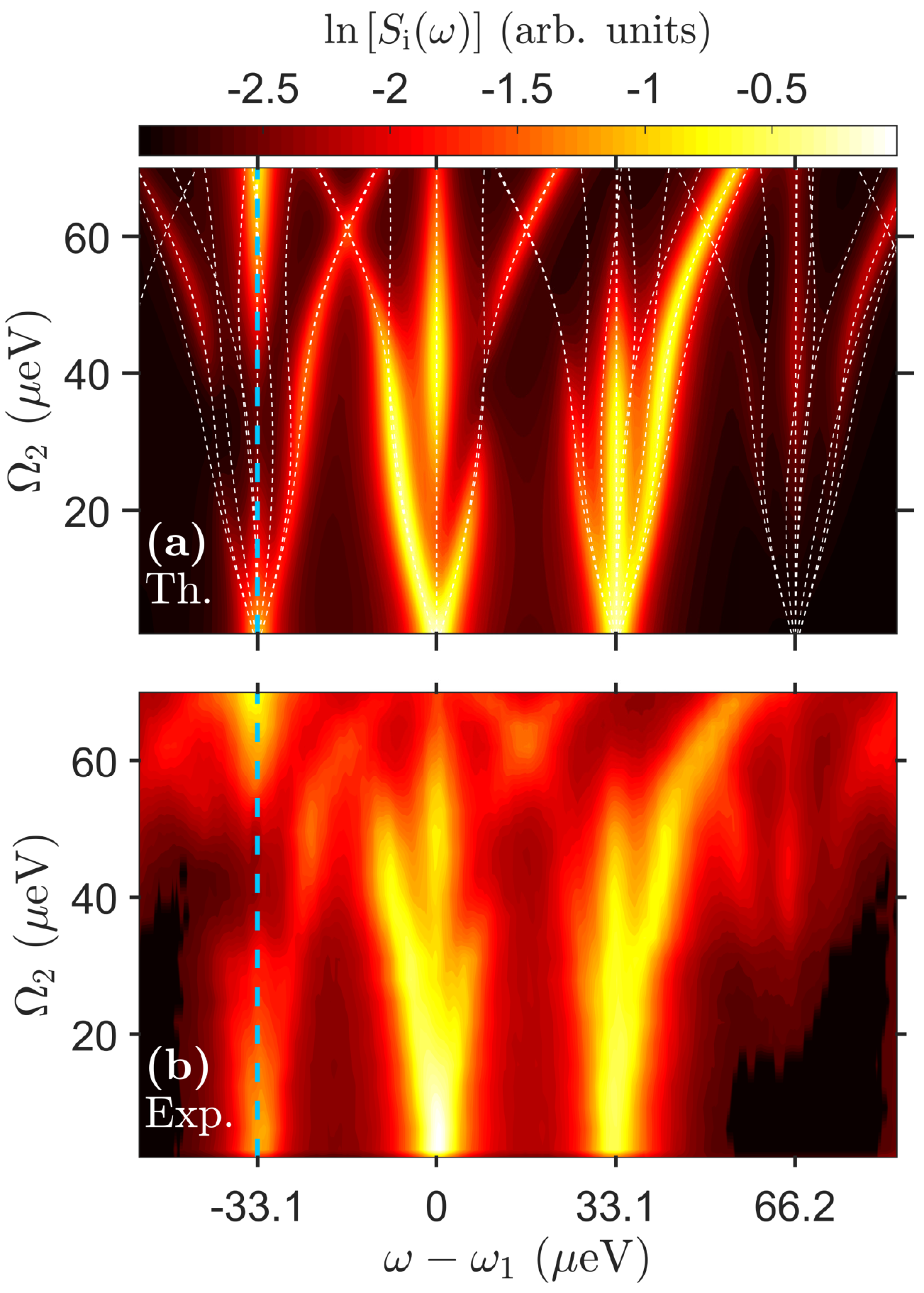}
    \caption{ \small  Theoretical and experimental log-scale spectra of a QD dressed by a primary laser with strength $\Omega_1 = 31.6 \ \mu \text{eV}$ and (red) detuning $\Delta_1 = 10 \ \mu \text{eV}$, and a second laser with varied power dressing the red sideband ($\Delta = -\sqrt{\Omega_1^2+\Delta_1^2}$). The locations of potential Floquet transitions are faintly overlayed in (a) as dashed white lines, and we use $\gamma = 1.66 \ \mu \rm{eV}$, and $\gamma' = 2 \ \mu \rm{eV}$. The location of the second laser at $\omega_2$ is shown as a blue vertical dashed line. }\label{fig4}
\end{figure}

Figure \ref{fig2} shows the theoretical and experimental emission spectra for the same scenario of one resonant laser plus a second with frequency resonant to the Rabi sidepeak induced by the first laser. Here both  a red-detuned and a blue-detuned second laser are considered.  Note that, in the theoretical spectrum, flipping the sign of $\Delta$ is formally equivalent to a sign change in $\omega-\omega_1$ (mirroring the spectrum) for this master equation; by considering Eqs.~\eqref{eq:me} and~\eqref{eq:s} and using the quantum regression theorem, it can be shown that $\Delta \rightarrow -\Delta$ is equivalent to $\langle \sigma_{\delta}^+(t)\sigma_{\delta}^-(t+\tau)\rangle \rightarrow \langle \sigma_{\delta}^+(t)\sigma_{\delta}^-(t+\tau)\rangle^*$, which has the same effect in Eq.~\eqref{eq:s} as taking $\omega - \omega_1 \rightarrow -(\omega - \omega_1)$.  We observe, in both theory and experiment, the disappearance and subsequent reappearance of the center (at $\omega=\omega_1$) resonance fluorescence transition line as a function of secondary drive power $\Omega_2$ --- a higher order effect in $\alpha_c$. \blue{We also observe the disappearance and subsequent reappearance of the spectral line at $\omega=\omega_2$, with increasing $\alpha_c$.} The additional triplets centered at plus (minus) twice the Rabi energy $\Omega_1$ for a blue (red) detuned second laser resulting from higher order effects are also clearly visible in the experimental and theoretical spectra. They are most pronounced for equal Rabi energies of both lasers. Note that we have made no specific effort to fit the decay and pure dephasing rates $\gamma, \gamma'$ to match experiment. Anomalously, we observe a crossing of spectral lines at $\omega-\omega_1 \approx 15 \ \mu \text{eV}$ in (c), and $\omega-\omega_1 \approx -15 \ \mu \text{eV}$ in (d), which correspond to Floquet transitions in our models, but are not reproduced by the full calculations.

In Fig.~\ref{fig3}(a,c), we plot the theoretical and experimental emission spectra for fixed drive strengths, with the first laser resonant and varying detuning of the second laser. In agreement with measurements by He {\it et al.},  \cite{he15} we observe suppression of the central transition when the second laser is in resonance with the sideband of the Mollow triplet at $\Delta=35~ \mu {\rm eV}$ or one of its subharmonics $\Omega_1/ \blue{m}$ for integer \blue{$m$} up to 3. This process can be associated with destructive quantum interference for the case where the second laser is resonant with the sideband, or a multiphoton quantum interference for the resonance with a subharmonic \cite{he15, ficek96,rudolph98}. Here, the second laser can couple via $m$ photons to the system. In addition to  suppression of spectral transitions, a series of triplets evolves separated by $\Omega_1/\blue{m}$. For the first subharmonic $\blue{m}=2$ at $\Delta \approx 17.5~\mu {\rm eV}$, a doublet at the center, a triplet at the low energy side and two at higher energies are visible. The second subharmonic resonance $\blue{m}=3$ at $\Delta\approx 11.7~\mu {\rm eV}$ is not as pronounced, but a similar spectrum is observable. In the limit $\Delta=\Delta_2=0$, besides their drive strengths, both lasers are only distinguishable by their relative phase, resulting in a modified Mollow triplet with plateau-like sidebands \cite{he15}.

To investigate a wider regime of the bichromatically excited QD, we also look at the case of a sideband dressed by the second laser again, while the first laser remains resonant with the exciton transition (Fig.~\ref{fig3}(b,d)). With an initially smaller Rabi energy of $\Omega_1=15~ \mu {\rm eV}$, we allow the drive strength to increase into the regime where $\alpha_c \gg1$. For the case where both lasers have similar strength $\Omega_1 \approx \Omega_2$ the same features as in Fig.~\ref{fig2} are observable. As a wider range of the spectrum is analysed here, triplets at higher multiples of the Rabi energy $\Omega_1 \times m$ are visible in both theoretical and experimental spectra. They are centered at $m=-2, 2, 3$. So far, only one emission triplet at twice the Rabi energy has been observed, in a bichromatically driven atom~\cite{yu97}.
For stronger drive strength of the second laser, the transition back to the Mollow excitation regime can be seen, where a dominant triplet begins to form centered around the second laser frequency $\omega_2$; the sidebands shift linearly away from $\omega = \omega_2$ with increasing drive strength $\Omega_2$.

In Fig.~\ref{fig4}, we plot the theoretical and experimental emission spectra for a detuned primary laser ($\Delta_1= 10 \ \mu \text{eV}$), where again the second laser dresses the sidepeak. As expected, with increasing drive strength of the second laser, similar spectra compared to the resonant scenario are observed, \blue{including the disappearance and subsequent reappearance of spectral peaks resonant with the laser frequencies,} while the asymmetry for small $\alpha_c$ is increased. In contrast to the single drive off-resonant Mollow triplet, which is symmetrical in the absence of pure dephasing, there is a strong inherent asymmetry in the Mollow-like spectrum for small but nonzero $\alpha_c$, which persists in our simulations even with $\gamma'=0$ (not shown)~\cite{gustin18}.
While dephasing generically broadens spectral peaks, it will also in general affect the relative spectral weights of differing peaks; pure dephasing in a bare-state basis appears as non-radiative dissipation when viewed in the system (Floquet) eigenbasis, which can violate the principle of detailed balance and thus generically changes the distribution of spectral weights (peak areas)~\cite{gustin18}. Specifically, for the regimes studied in this work, we do find in our simulations that dephasing will slightly change the distribution of spectral weights, but this effect is generally insignificant to qualitatively affect our plots, and largely serves to broaden the spectral linewidths.

\section{Conclusions}\label{sec:c}
In conclusion, we have presented high resolution spectroscopy measurements of a single QD dressed simultaneously by two coherent laser drives in general non-perturbative regimes with respect to the laser drive strengths and detunings. These measurements are reproduced with excellent accuracy by a time-dependent Lindblad master equation model which can be easily generalized to incorporate different environmental couplings (e.g., cavity, phonon scattering). The transition lines are identifiable using semi-analytic Floquet theory, with higher order harmonics required for accurate location of the spectral lines, which is indicative of the regime with two strong components of the bichromatic field. 

For the excitation scenario of one laser resonant with the QD, and one resonant to a sidepeak of the resulting Mollow triplet, further evidence of the regime of excitation with two strong fields can be seen in the formation of additional spectral triplets at two and three times the central laser Rabi energy, the disappearance and subsequent re-emergence of the central spectral peak \blue{as well as the peak resonant with the secondary laser}, and the transition to a detuned Mollow triplet centered around the second laser as its drive strength is increased well beyond the first laser strength. 

We also have considered a similar excitation scenario, but with both lasers off-resonant, and observed an enhanced inherent spectral asymmetry relative to the resonant primary laser case, in contrast to the monochromatic case where the spectrum is completely symmetric in the absence of additional dephasing mechanisms. All of these features are very well reproduced by theoretical calculations. These results reveal broad potential for spectral \blue{and density of optical states} engineering using Floquet Hamiltonians (multi-color coherent excitation) which is highly general and rich in structure and optical physics. 

\blue{A natural extension of this work could be to study three-color (trichromatic) driving, either by three coherent lasers or use of an amplitude-modulated drive (which is equivalent to a bichromatic drive), where the phase coherence between different drives has an effect on the emission spectra, and could be (for example) used to destroy or restore spectral symmetry.}

\acknowledgements
We acknowledge funding from Queen's University, the Canadian Foundation for Innovation, the Natural Sciences and Engineering Research Council of Canada, the European Union’s Horizon 2020 research and innovation programme under grant agreement No. 820423 (S2QUIP), the German Federal Ministry of Education and Research via the funding program Photonics Research Germany (contract number 13N14846) and the Deutsche Forschungsgemeinschaft (DFG, German Research Foundation) under Germany's Excellence Strategy – EXC-2111 – 390814868. KM acknowledges support from the Bavarian Academy of Sciences and Humanities. SH acknowledges financial support from the Alexander von Humboldt Foundation.



\appendix
\section{Floquet Theory Analysis}\label{A:A}

\blue{In this appendix, we show how to calculate the Floquet energy states of the system, which give the manifold of potential optical transitions seen in the spectrum (see Ref.~\onlinecite{tannor})}. By exploiting the discrete time-translational symmetry of the Hamiltonian $H(t+ nT) = H(t)$ for $n \in \mathbb{Z}$, where $T = 2\pi/|\Delta|$, we construct a simple unitary Floquet model which neglects dissipation to identify the resonances of the system. 

For the general time-dependent problem $i\frac{\rm{d}}{\rm{dt}}\ket{\psi} = H(t)\ket{\psi}$, the solution is given formally by $\ket{\psi(t)} = U(t,t_0)\ket{\psi(t_0)}$, where $U(t,t_0) = \hat{\tau}\exp{\left(-i \int_{t_0}^t H(t')dt'\right)}$ and $\hat{\tau}$ is the time-ordering operator. The essential utility of Floquet theory is the transformation of this time-dependent problem to a time-independent one given by an infinite matrix deduced by Fourier expansion of $\ket{\psi(t)}$. This result is possible by Floquet's theorem~\cite{shirley65,breuer}, which states that there exists a complete set of states $\ket{\psi_\lambda(t)}$ indexed by $\lambda$ that satisfy
\begin{equation}
    \ket{\psi_\lambda(t)} = e^{-i \epsilon_\lambda t} \ket{\phi_\lambda(t)},
\end{equation}
where $\epsilon_\lambda$ denotes a real Floquet quasienergy, and $\ket{\phi_\lambda(t)}$ has periodicity $T$. 

This immediately yields the eigenvalue problem:
\begin{equation}\label{eigen}
H_F\ket{\phi_\lambda} = \epsilon_{\lambda}\ket{\phi_\lambda},
\end{equation}
with the Floquet Hamiltonian operator $H_F = H(t) - i \frac{\rm{d}}{\rm{dt}}$. Although time-dependent, the Floquet states $\ket{\phi_\lambda(t)}$ form a complete basis for any value of $t$, and as such the general solution to the Schr{\"o}dinger equation can be given as
\begin{equation}\label{eq:fullsol}
    \ket{\psi(t)} = \sum_{\lambda}c_\lambda e^{-i\epsilon_\lambda t }\ket{\phi_\lambda(t)},
\end{equation}
where the $c_{\lambda}$ are time-independent complex coefficients. As the states $\ket{\phi_\lambda}$ are periodic, Eq.~\eqref{eq:fullsol} suggests that energies in Floquet systems are only conserved modulo $\Delta$, and it can be shown the transition resonances of the system occur at differences between Floquet energies~\cite{shirley65,breuer}.

For our two-level model, we can expand $\ket{\phi_\lambda} = c_{g,\lambda}(t)\ket{g} + c_{x,\lambda}(t)\ket{x}$, where $c_{g,\lambda}(t)$ and $c_{x,\lambda}(t)$ are periodic with period $T$. We expand them as Fourier series,  $c(t) = \sum\limits_{m=-\infty}^{\infty} \! c^{(m)}e^{i m \Delta t}$, where $m$ is an integer, and insert the result into Eq.~\eqref{eigen}:
\begin{equation}
\sum\limits_{m=-\infty}^{\infty}\sum\limits_{\beta=g,x}\big(H_F - \epsilon_\lambda\big) c_{\beta,\lambda}^{(m)}e^{i m \Delta t}\ket{\beta} = 0.
\end{equation}
Taking the inner product with $\bra{\alpha}$, 
\begin{equation}
\sum\limits_{m,\beta}\Big[H_{\alpha,\beta}\! +\! m\Delta\delta_{\alpha,\beta}\Big]c_{\beta,\lambda}^{(m)}e^{i m \Delta t}\! =\! \sum\limits_{m}\epsilon_\lambda c_{\alpha,\lambda}^{(m)}e^{i m \Delta t},
\end{equation}
with $H_{\alpha,\beta} = H_{\alpha,\beta}(t)=\bra{\alpha}H(t)\ket{\beta}$. Multiplying by $e^{-i n \Delta t}/T$, where $n$ is an integer, and integrating from $0$ to $T$, then
\begin{align}\label{eigen2}
\sum\limits_{m,\beta}\!\Big[\frac{1}{T}\!\int_{0}^{T}\!\!\!dt H_{\alpha,\beta} e^{i(m\!-\!n) \Delta t}\! +\! n \Delta\delta_{\alpha,\beta} \delta_{m,n}\Big]\!c_{\beta,\lambda}^{(m)}\!  =\! \epsilon_\lambda c_{\alpha,\lambda}^{(n)}.
\end{align}

Equation~\eqref{eigen2} is an eigenvalue equation for the eigenvector of Fourier coefficients $c_{\alpha,\lambda}^{(n)}$ and eigenvalue $\epsilon_{\lambda}$, with matrix elements specified by the row indexed by $(\alpha,n)$ and column indexed by $(\beta,m)$, and equal to the quantity in square brackets---the matrix representation of $H_F$. This matrix can be realized computationally by truncating the number of integers $n,m$ considered, and letting each combination of $(n,m)$ correspond to a 2-by-2 block of elements corresponding to the matrix elements of the two-level Hilbert space Hamiltonian, plus $n \Delta$ on the diagonal elements. 

To extract the frequencies of the spectral lines that show up in the emitted spectrum, the eigenvalues of Eq.~\eqref{eigen2} can be found numerically for a given truncation of integers $n,m$, and the potential transitions are given by the Floquet-dressed resonant frequencies $\tilde{\omega}_{\lambda,\lambda'} = \omega_1 + (\epsilon_\lambda-\epsilon_{\lambda'})$, where $\lambda \in \{1,2,...,M\}$, and $M=2(2N+1)$. Here, $N$ is the order of the harmonics considered in the Floquet matrix (i.e. $n,m = 0, \pm 1, ..., \pm N$). Thus, considering up to order $N$ yields $M^2$ potential transitions, although not all these transitions need be driven, and thus may not all show up in the spectrum. Many will also be degenerate. Increasing the Floquet order increases the accuracy of the locations of the spectral lines, as well as increasing the amount of spectral resonances that can be identified.

We now show explicitly how the eigenvalues can be extracted from the matrix representation of $H_F$ to a certain harmonic order. Note that as the time-dependent component of the Hamiltonian becomes appreciable relative to the other timescales of the system (for the Mollow regime, this corresponds to the magnitude of $\alpha_c$), higher harmonic orders must be included in the matrix for accurate results. 

As an example, if we consider transitions up to order $N=1$ in the harmonic expansion, we can represent $H_F$ from Eq.~\eqref{eigen2} in the basis $(1,0,-1)$:
\begin{equation}
H_F = 
\begin{bmatrix}
    \mathbf{M}_{1,1} & \mathbf{M}_{1,0} & \mathbf{0} \\
    \mathbf{M}_{1,0}^{\dagger} & \mathbf{M}_{0,0} & \mathbf{M}_{1,0} \\
    \mathbf{0} & \mathbf{M}_{1,0}^\dagger & \mathbf{M}_{-1,-1} 
\end{bmatrix},
\end{equation}
where we let bold denote 2-by-2 matrices in the two-level system basis $(x,g)$, and we have made use of the fact that, clearly, $\mathbf{M}_{n,m} = \mathbf{M}_{m,n}^{\dagger}$, and $\mathbf{M}_{m,m \mp 1}=\mathbf{M}_{m \pm 1,m}$. Furthermore, elements with $|n-m|>1$ vanish due to the form of the Hamiltonian in Eq.~\eqref{hamiltonian}. Evaluating the matrix elements explicitly, to order $N=1$, then
\begin{align}
H_F &= \frac{1}{T}\int_0^{T}dt
\begin{bmatrix}
    \mathbf{H}+\Delta\mathbf{1} & \mathbf{H}e^{-i\Delta t} & \mathbf{0} \\
    \mathbf{H}e^{i\Delta t} & \mathbf{H} & \mathbf{H}e^{-i\Delta t} \\
    \mathbf{0} & \mathbf{H}e^{i\Delta t} & \mathbf{H} - \Delta\mathbf{1} 
\end{bmatrix} 
\nonumber \\ \nonumber \\
&= \frac{1}{2}
\begin{bmatrix}
    2(\Delta_1\! +\! \Delta) & \Omega_1 & 0 & 0 & 0 & 0 \\
    \Omega_1 & 2\Delta & \Omega_2 & 0 & 0 & 0 \\
     0 & \Omega_2 & 2\Delta_1 & \Omega_1 & 0 & 0 \\
     0 & 0 & \Omega_1 & 0 & \Omega_2 & 0 \\
     0 & 0 & 0 & \Omega_2 & 2(\Delta_1 \!-\! \Delta) & \Omega_1 \\
     0 & 0 & 0 & 0 & \Omega_1 & -2\Delta
\end{bmatrix},
\end{align}
the eigenvalues of which give the energy spectrum of the field-dressed Floquet states in the frame of the laser with frequency $\omega_1$ (to harmonic order $N=1$). 

\begin{figure}[thb]
        \includegraphics[width = 0.95\columnwidth]{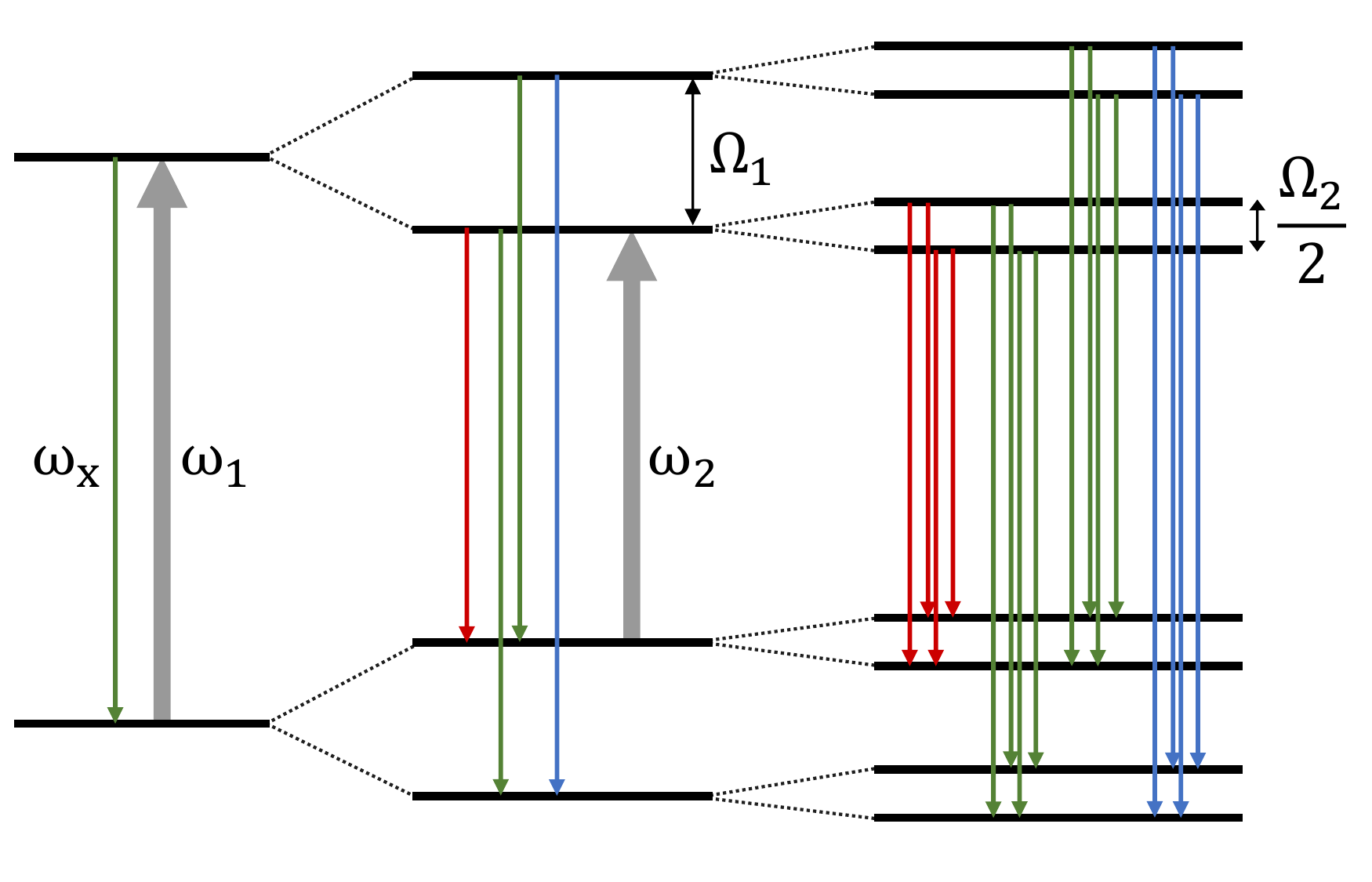}
    \caption{\blue{Simplified schematic of relevant doubly-dressed QD energy levels for the case of resonant excitation with a strong primary drive. Exciting the exciton transition resonantly with a laser of Rabi energy $\Omega_1$ and frequency $\omega_1=\omega_x$ (left) results in a coupled system of laser and two-level system with a manifold of doublet states split by $\Omega_1$ and separated by $\omega_1$ (middle). The Mollow triplet is formed in this singly dressed system by the transitions shown in blue, green (degenerate) and red lines. Introducing a weaker second laser with Rabi energy $\Omega_2$, here resonant with the sideband at lower energy, each peak of the singly dressed Mollow triplet itself splits into a new triplet, with peaks separated by $\Omega_2/2$ (right), with the central transition at $\omega_1$ suppressed due to quantum interference. The second laser is assumed to be much weaker that the first ($\alpha_c \ll 1$), such that the relevant spectral lines from the manifold of Floquet states can be identified by considering only two manifolds, separated by $\omega_1$, as shown on the right. For larger $\alpha_c$, additional triplets appear at higher integer multiples of the singly dressed system splitting, the ninth center peak reappears, and the location of each of the peaks changes due to effects associated with higher order Floquet states and the breakdown of perturbation theory. 
    } \
    }
    \label{fig_diagram}
\end{figure}

\blue{The eigenvalues of this matrix are found from the characteristic equation:
\begin{align}\label{eq:ch}
    g(\tilde{\omega})\Bigg[ & f(\tilde{\omega})\left(1 - 4(\tilde{\Delta}-\tilde{\omega})(\tilde{\Delta}_1+\tilde{\Delta}-\tilde{\omega})\right) \nonumber \\ & +4\alpha_c^2(\beta(\tilde{\omega})+\tilde{\omega})(\tilde{\Delta}_1+\tilde{\Delta}-\tilde{\omega})\Bigg]=0,
\end{align}
where 
\begin{equation}
    g(\tilde{\omega}) = (1+\tilde{\Delta}_1^2) -4 (\tilde{\omega} + \tilde{\Delta} - \frac{1}{2}\tilde{\Delta}_1)^2, 
\end{equation}
\begin{equation}
    f(\tilde{\omega}) = 1 + 4(\beta(\tilde{\omega})+\tilde{\omega})(\tilde{\Delta}_1 - \tilde{\omega}),
\end{equation}
and
\begin{equation}
    \beta(\tilde{\omega}) = \alpha_c^2 \frac{(\tilde{\omega}+\tilde{\Delta})}{g(\tilde{\omega})},
\end{equation}
with $\tilde{\omega} = \omega/\Omega_1$, $\tilde{\Delta}_1 = \Delta_1/\Omega_1$, and $\tilde{\Delta} = \Delta/\Omega_1$. To this order, two of the energy levels are independent of $\alpha_c$ and can be found immediately as $\frac{1}{2}\tilde{\Delta}_1 - \tilde{\Delta} \pm \frac{1}{2}\sqrt{1+\tilde{\Delta}_1^2}$--- from which it follows why the location of transitions separated by integer multiples of $\sqrt{\Omega_1^2+\Delta_1^2}$ remain unaffected by the second laser drive strength when $\alpha_c \ll 1$. 

While exact solutions are not typically available for the other four energy values, one can use perturbation theory to calculate the Floquet energies in certain regimes. For example, for $\alpha_c = \tilde{\Delta}_1=0$, we have the eigenvalues $\{\tilde{\omega}\}= \{ \pm 1/2, \tilde{\Delta} \pm 1/2, -\tilde{\Delta} \pm 1/2\}$. To compare with the full calculation in Fig.~\ref{fig1}, we can take $\tilde{\Delta} = -1$, and seek a perturbation expansion in $\alpha_c \ll 1$. We find a shift of energy levels $\{\tilde{\omega}\}= \{ \frac{1}{2} + \frac{3}{64}\alpha_c^2 \pm \frac{1}{4}\alpha_c,- \frac{1}{2} - \frac{3}{64}\alpha_c^2 \pm \frac{1}{4} \alpha_c,\pm (\frac{3}{2} + \frac{3}{32} \alpha_c^2)\} + \mathcal{O}(\alpha_c^3)$, which agrees with the $N=1$ transition lines of Fig.~\ref{fig1}(a) up to roughly $\Omega_2 \approx 20 \ \mu \text{eV}$.
This perturbative regime corresponds to the simple explanation in the main text of ``doubly-dressed" states, where the second laser further splits the singly-dressed laser-QD states, as well as a Bloch-Siegert-like shift due to the $\alpha_c^2$ terms. Neglecting the highest and lowest energy states (which predominantly effect the higher order harmonic peaks at $\omega - \omega_1 \approx \pm  \Omega_1$), the splitting in the Mollow triplet's three peaks is (to order $\alpha_c$) $\pm \frac{\Omega_2}{2}$, in agreement with previously known results~\cite{ficek96,he15}. \blue{In Fig.~\ref{fig_diagram} we provide a simplified energy level diagram for the case of resonant excitation}. However, note that this result is only valid in this perturbation regime $\alpha_c \ll 1$; for larger second laser strengths, in addition to the analytic perturbative solution breaking down, the higher Floquet harmonics are furthermore clearly needed to describe the dynamics accurately. Indeed, even for very small $\alpha_c$, the spectral lines at $\omega - \omega_1 \approx \pm  \Omega_1$ require higher order Floquet harmonics for accurate characterization. Thus, the regimes studied in this work require generic numerical techniques.

We can also consider the case where the primary laser is detuned from the two-level system resonance, and the second laser again dresses a sideband. As an example, we study the configuration shown in Fig.~\ref{fig4}, where we have the detuning $\tilde{\Delta} = -\sqrt{1+\tilde{\Delta}_1^2}$, and $\tilde{\Delta}_1 > 0$. The energy levels for $\alpha_c=0$ are $\tilde{\omega}^{(m)}_0 = \frac{1}{2}\tilde{\Delta}_1 +\frac{2m-3}{2}\sqrt{1+\tilde{\Delta}_1^2}$, $m = 0,1,2,3$. Once again we seek a perturbative solution to Eq.~\eqref{eq:ch} in terms of $\alpha_c$. To order $\alpha_c$, we find $\tilde{\omega}^{(0)} = \tilde{\omega}^{(0)}_0$, $\tilde{\omega}^{(3)} = \tilde{\omega}^{(3)}_0$, and we have a degeneracy splitting in $\tilde{\omega}^{(1)}$ and $\tilde{\omega}^{(2)}$:
\begin{align}
  &\tilde{\omega}^{(1)}_0 \rightarrow \tilde{\omega}^{(1)}_{\pm} =   \tilde{\omega}^{(1)}_0 \pm \frac{1}{4}\eta \alpha_c \\ &
        \tilde{\omega}^{(2)}_0 \rightarrow \tilde{\omega}^{(2)}_{\pm} =   \tilde{\omega}^{(2)}_0 \pm \frac{1}{4}\eta \alpha_c,
\end{align}
where
\begin{equation}\label{eq:eta}
\eta = 1 - \frac{\Delta_1}{\sqrt{\Omega_1^2+\Delta_1^2}}.
\end{equation}
Here, the splitting from each (singly dressed) peak of the off-resonant Mollow triplet is $\pm \frac{\eta \Omega_2}{2}$ for small $\alpha_c$, which agrees with experiment and full simulation in Fig.~\ref{fig4} ($\eta \approx 0.70$), as well as the fully quantum mechanical dressed state model from Ref.~\cite{ficek96}. 
}

\blue{\section{Quantum Dot Characterization}\label{A:B}

\begin{figure*}[thb]
        \includegraphics{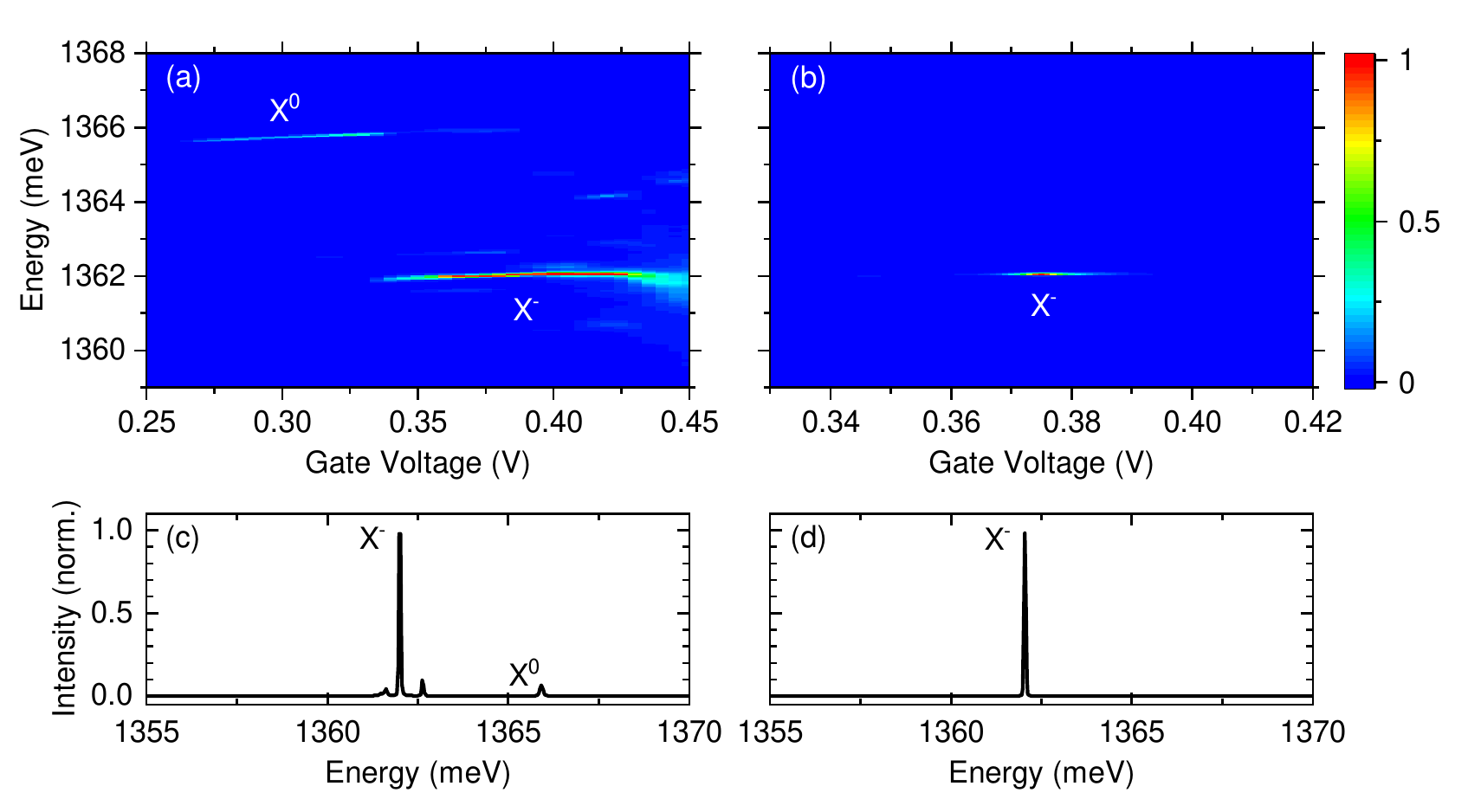}
    \caption{\blue{\small (a) Voltage-dependent photoluminescence series of a QD excited in the wetting layer with a 850 nm ($\approx 1.46$ eV) laser diode. The observed charge plateaus correspond to the neutral and negatively charged exciton transition. The emission spectrum at 0.375 V is shown below (c). Resonant excitation (b,d) of the trion transition suppresses other decay channels, thus a clean emission spectrum with a single peak is observed.}}
    \label{PLV}
\end{figure*}

To study the excitonic transitions of the single QD, we excite the system in the wetting layer with a continuous-wave laser diode with 850 nm wavelength. Generated charge carriers can be trapped by the QD and thermalize non-radiatively to the energetically lowest state before recombining by emission of a single photon. The fabricated Schottky diode structure leads to a built-in electric field which can be additionally controlled by applying an external voltage. Thus, we can adjust the bending of the band structure and therefore the relative difference between the Fermi level of the n-doped GaAs layer and the quantized states of the QDs. 

Figure~\ref{PLV}(a) shows photoluminescence measurements of the quantum dot as a function of the applied voltage. For low external voltage (Fig.~\ref{PLV}(a) left), the built-in field is strong enough for all charge carriers to tunnel to the contacts before recombining radiatively, as their decay rate is larger than the tunneling rate. For increasing voltage the bands flatten, making it possible to trap excitons in the quantum dot, which decay radiatively. Increasing the voltage further, a sharp crossover to a new emission line can be observed. In this regime the band alignment allows tunneling of an electron from the n-doped layer into the quantum dot. Thus we observe luminescence from the negatively charged trion transition as is shown in Fig.~\ref{PLV}(c) which is shifted in emission energy due to the Coulomb interaction.
Resonant excitation of the two-level system (trion transition) suppresses the generation of excess charge carriers in the wetting layer, resulting in one clean and sharp emission line (Fig.~\ref{PLV}(d)). Tuning the voltage (Fig.~\ref{PLV}(b)) allows for fine-tuning of the emission frequency via the quantum-confined Stark effect, which allows us to bring the excitation laser and transition frequency into exact resonance.

\begin{figure}[thb]
        \includegraphics{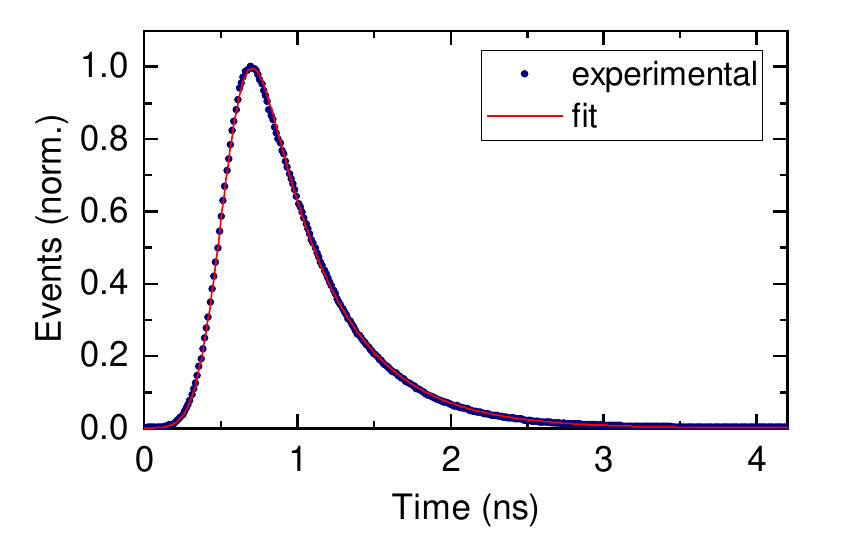}
    \caption{\blue{Measured radiative decay of the negatively charged exciton state. Fitting the data (red solid line) yields a lifetime of $455$ ps.}}
    \label{lifetime}
\end{figure}

In addition, we perform a lifetime measurement of the single quantum dot to investigate the radiative decay rate of the negatively charged trion. For that purpose, the system is excited with a short resonant laser pulse and the emission is recorded time-resolved as a histogram (Fig.~\ref{lifetime}). The decay describes spontaneous emission as a function of time. By fitting the data while taking the instrument-response function into account, we obtain a value for the lifetime of the excited state of $T_1 = 1/\gamma = 455$ ps.

}
\section{Resonance Fluorescence under a Single Strong Coherent Drive}\label{A:C}

\begin{figure}[thb]
        \includegraphics[width = 0.95\columnwidth]{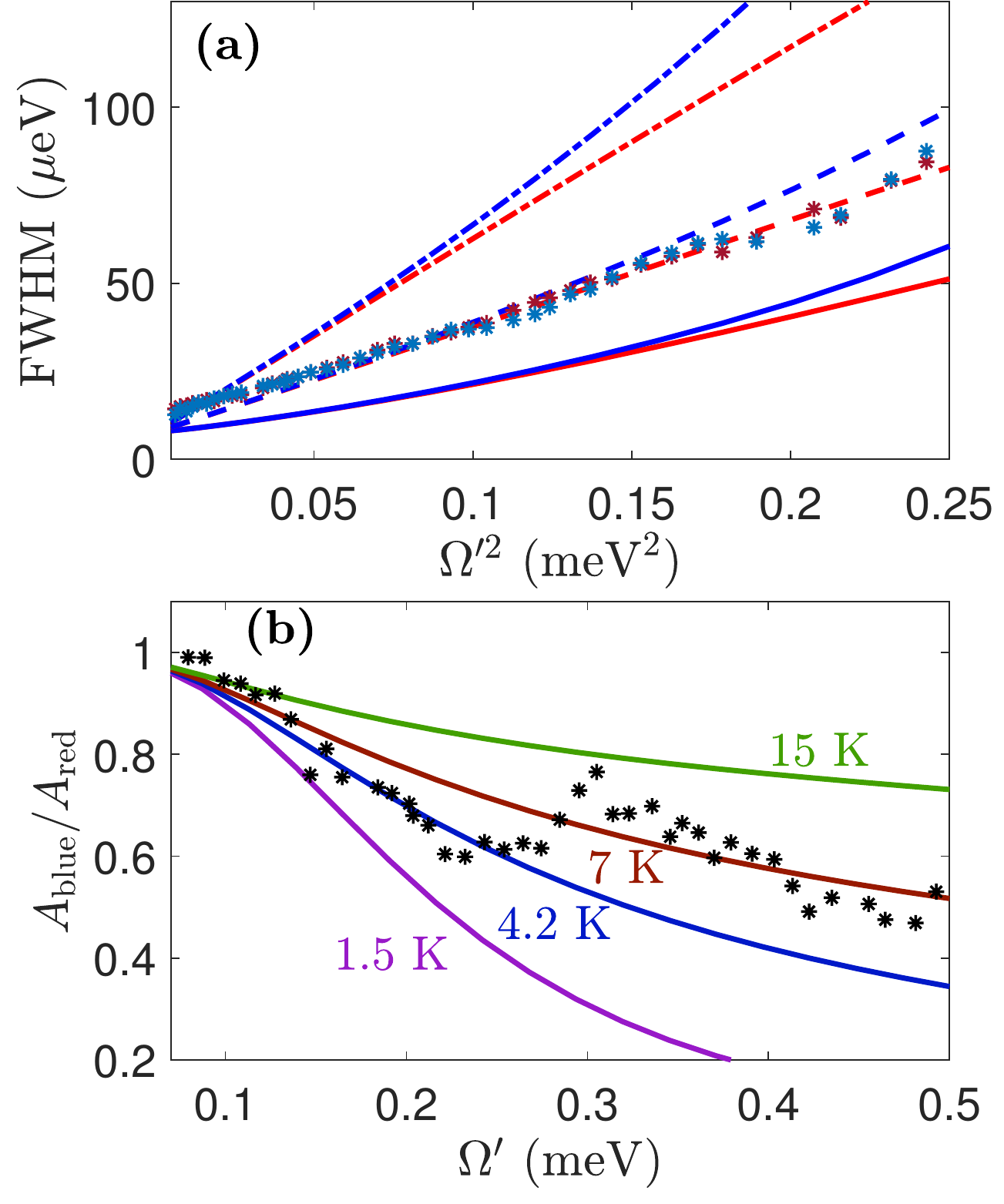}
    \caption{ \small (a) Fitted FWHM of red (shown as red) and blue (shown as blue) Mollow sidepeaks for experimental data (points), and theoretical curves at temperatures $T=7$ K (dash-dotted lines), $T =4.2$ K (dashed lines), and $T=1.5$ K (solid lines). (b) Ratio of the integrated intensities of blue and red detuned sidebands as a function of the Rabi energy. Experimental data shown as black points, and theoretical curves are given for various temperatures. }
    \label{fig_A}
\end{figure}

To determine the regime in which pump-induced electron-phonon interactions become significant, we study the resonance fluorescence spectrum of a QD driven by a single resonant coherent drive, where the drive strengths are large enough to enter a regime in which electron-phonon scattering is appreciable.  For theoretical calculations, we use the polaron transform model from Ref.~\cite{2gustin18}, with a time-independent resonant drive on a two-level system without a cavity mode. The phonon scattering is characterized by the phonon spectral function $J(\omega) = \alpha\omega^3 \exp{\left[-\frac{\omega^2}{2\omega_b^2}\right]}$, with phonon coupling strength $\alpha$ and phonon cutoff frequency $\omega_b$. We let $\omega_b = 0.9 \ \text{meV}$, consistent with previous experiments~\cite{weiler12}. By curve fitting the data of Fig.~\ref{fig_A} (described below) with reference to Eq.~\eqref{eq:naz}, we find parameters $\alpha = 0.1 \ \text{ps}^2$ and $\gamma'= 10 \ \mu$eV; Eq.~\eqref{eq:s} remains valid for calculation of the incoherent spectrum, but the $t$-integral becomes trivial in the steady state condition. We use a decay rate of $\gamma = 1.66 \ \mu \text{eV}$. Note that here we include the full phonon sideband that arises from the polaron solution~\cite{roy12}. To account for heating of the QD sample induced by the large drive strengths, which results in an exciton resonant frequency shift in the experimental data, we sweep the laser detuning to find the resonant condition for each of the power-dependent measurements.

Under resonant excitation, the incoherent resonance fluorescence spectrum exhibits the well-known Mollow triplet shape, and by curve fitting to Lorentzian functions (for $\Omega' \gg \gamma, \gamma'$, where $\Omega'$ denotes the polaron-renormalized Rabi energy~\cite{2gustin18}, such that the peaks are separated and well-represented by Lorentzians), we extract values of the full width(s) at half maximum (FWHM) and spectral weight.  In Fig.~\ref{fig_A}(a), we plot the fitted FWHM for the Mollow sidepeaks as a function of the square of the drive amplitude. The parameters $\alpha = 0.1 \ \text{ps}^2$ and $\gamma' = 10 \ \mu\text{eV}$ were found by curve fitting to Eq.~\eqref{eq:naz} (with $\Omega \rightarrow \Omega'$) with an additional offset determined by $\gamma'$; note that here we are simply using $\gamma'$ as a crude substitute for peak broadening that occurs over long timescales --- we use a much smaller value in the main text as pure dephasing in the general case should correspond to processes that occur on the timescales of the excitation dynamics. As one can not rule out other dephasing processes which scale with the laser power, more precisely, 
these measurements suggest $\alpha \lesssim 0.1 \  \text{ps}^2$. Indeed the difference in the theoretical red and blue curves (which is minimal in the experimental data) is much less visible for smaller values of $\alpha$ (e.g., $\alpha = 0.06 \ \text{ps}^2$, consistent with Ref.~\cite{weiler12}), indicating this is likely the case.

In Fig.~\ref{fig_A}(b), we plot the ratio of red and blue Mollow sidebands as a function of drive amplitude. Our simulations (not shown) reveal that the sideband ratio is nearly independent of phonon parameters $\alpha$ and $\omega_b$, with the vast majority of its dependency coming from the temperature. 
This is understood by recalling that temperature is what determines the amount of phonons available in the thermal bath, which is the origin of this asymmetry. 
Making a direct comparison of experiment and theory is complicated because of the bump in the experimental data at $\Omega' \approx 0.3$ meV, which may be due to confined phonon effects~\cite{weiler12}, and because the high laser powers used in these single-drive experiments are large enough to induce heating in the sample, resulting in a power-dependent temperature. However, by analysis of the slope of the data, the effective temperatures can be constrained to $T \lesssim 7$ K, with the theoretical curve at $T=4.2$ K being in excellent agreement with experimental data for small values of $\Omega'$. Note that if we use $\alpha \sim 0.05 \ \text{ps}^2$, the experimental data of the ratio of the sidepeaks as a function of drive strength lines up much more closely with the $T=1.5$ K curve, indicating that $0.05 \ \text{ps}^2 \lesssim \alpha \lesssim 0.1 \ \text{ps}^2$, in accordance with our discussion of Fig.~\ref{fig_A}(a).
Note that we have not fitted the phonon cutoff frequency $\omega_b$, but this parameter does not enter into the low-drive dephasing rate in Eq.~\eqref{eq:naz} and thus has a smaller influence in this regime, and as well it is constrained by the size of the QD. 

Comparison of theory and experiment for both these plots suggests that a temperature of between $T= 4.2$ K and $T = 7$ K gives the closest agreement between the two results. Although in absence of a drive the sample is measured to be at $T = 4.2$ K, the strong laser powers in these measurements cause a power-dependent heating which is not observably present for the bichromatic measurements. Thus, these results suggest that a polaron transform phonon model at $T = 4.2$ K, using $\alpha \lesssim 0.1 \ \text{ps}^2$ is very likely appropriate for the main results of this paper. As discussed in Sec.~\ref{sec:th}, these parameters are expected to give negligible phonon effects (beyond pure dephasing) for the strengths of the bichromatic drives used in this work, further justifying the phenomenological pure-dephasing treatment of the electron-phonon interaction. Note that without any phonon coupling in the simulations, the functions in these plots become trivially identically equal to one.

\bibliography{main_bichromaticre_final}

\end{document}